\documentclass[aps,prd,
nofootinbib,
reprint,
groupedaddress,amsmath,amssymb, showpacs
]{revtex4-1}

\usepackage{graphicx}
\usepackage{bm}
\usepackage{dcolumn}
\def \pbar{\bar{p}}
\begin{document}


\title{Gluon-gluon elastic scattering amplitude in classical color field of colliding protons}


\author{\textsc{Man-Fung Cheung}}
\email[]{mfcheung@physics.utexas.edu}
\author{\textsc{Charles B. Chiu}}
\email[]{chiu@physics.utexas.edu}
\affiliation{Center for Particles and Fields and Department of Physics\\
University of Texas at Austin, Austin, TX 78712, USA}


\date{\today}

\begin{abstract}
We present a formalism for gluon-gluon elastic scattering in the presence of the classical color field of the protons in high energy collision.  The classical field is obtained by solving the classical Yang-Mills equation in the covariant gauge and treated as a prescribed background for the quantum gluons involved in the scattering process.  The interaction between the classical field and the quantum gluon modifies the gluon propagator, and, in turn, the $gg\rightarrow gg$ amplitude.  The modified gluon propagator is derived to the first non-zero order of the classical field using the Gaussian approximation in Color Glass Condensate and shown to satisfy the generalized Slavnov-Taylor identity.  This formalism is the theoretical basis for our recently proposed classical color field modified minijet model where we show that the $pp$ and $\pbar p$ cross section data from $\sqrt{s}=5$ GeV to $30$ TeV can be satisfactorily fitted and the model predicts a $(\ln s)^2$ behavior for large $s$, which saturates the asymptotic behaviour of Froissart bound. 
\end{abstract}


\maketitle


\section{\label{sec:1}Introduction}
In this paper we present a formalism for gluon-gluon scattering in the presence of classical field due to the protons in high energy $pp$ collision. This formalism is the theoretical basis of our recently proposed classical field modified minijet model.  We recall that the QCD-inspired minijet model was first introduced in the 70's \cite{Cline1973}.  At the time, it was noticed that the rise of the total cross section was very similar to the jet production cross section.  In this context, it is natural to separate the total cross section into a soft component and a hard component.  The hard component is to be calculated through the pQCD motivated minijet model.  However, the minijet cross section rises too rapidly, namely, if one assume the gluon distribution as $xg \propto x^{-J}$ with $J>1$ for small $x$, the cross section $\sigma \sim s^{J-1} \ln s$, that violates the Froissart bound.  In order to restore unitarity, since then the minijet model has been incorporated in the eikonal model, usually referred to as eikonal minijet model, by various authors in an attempt to tame the rise and explain the data quantitatively \cite{Pancheri1984,Gaisser1985,
Durand1987,Capella1987,Godbole2005,Achilli2011}.  

We consider the modification of gluon-gluon scattering amplitude of the minijet model due to interaction between the gluons involved in the gluon-gluon elastic process and the medium consists of small-$x$ gluon.  The small-$x$ gluons are described by the classical effective theory calculated using the Gaussian approximation \cite{Iancu2002b} in Color Glass Condensate (CGC), while the scattering gluons see the small-$x$ gluon as a background field.  In the theory of CGC, one first defines a scale in the longitudinal momentum fraction, $x$, or rapidity, then assumes the collection of the large-$x$ partons as a classical random source $\rho$ which generates a classical field through Yang-Mills equation.  The classical field represents the field of the small-$x$ gluon.  A physical observable is an average over the configuration space of $\rho$ with a weight function $W[\rho]$ \cite{McLerran1994,McLerran1994a,McLerran1994b}.  When the scale $x$ decreases to $x'=x-dx$, the source includes the partons from $x=1$ down to $x=x'$.  The new source at scale $x'$ consists of a new layer of parton (mainly gluon) compared to the source at scale $x$.  This new layer of source comes from the quantum gluon fluctuation and further induces a correction in the classical field.  This corresponds to having a new weight function at the new scale $x'$.  The $x$-dependence of the weight function is governed by a renormalization group (RG) equation \cite{Ayala1995,Ayala1995b,JalilianMarian1997,Iancu2001}.  

In our approach, we apply the idea of describing the small-$x$ gluon field as a classical field and the physical observable is an average over configurations of the source.  However, both the classical field ans the source are prescribed inputs of our approach.  Once longitudinal momentum fractions, $(x_1,x_2)$, of the incident gluons involved in the scattering process are specified, the classical field and the course are fixed.  The classical field provides an arena to the scattering process.  The gluons in $gg \rightarrow gg$ Feynman diagram do not only interact with each other but also with the classical field inside the protons.  The longitudinal momentum fractions of the scattering gluons serves as the separation scale of the classical field and the source of their parent protons.  The scattering amplitude will depend on the classical source of the protons through the interaction with the classical field.  The physical amplitude is then obtained by taking the average over the random sources of both protons.  Although an analytic solution of the RG equation is difficult to obtain, an approximate solution which preserves the Gaussian structure is suggested in \cite{Iancu2002b} (IIM model).  The authors claimed that the approximation captures the relevant physics in both saturated and dilute regime:  namely, the two-point correlation of the source $\mu_x(x_\perp,y_\perp) = \int \langle \rho^a(x_\perp,x^-)\rho^a(y_\perp,y^-) \rangle dx^-dy^-$ obeys BFKL equation for small $(x_\perp-y_\perp)$ and demonstrates color neutrality at long range, $(x_\perp-y_\perp) \gg 1/Q_s(x)$.  The same Ansatz in \cite{Iancu2002b} will be used in this paper to characterize the classical field.  We first calculate the modification to the $gg$ amplitude up to the first non-zero order of the classical source and coupling constant.  Through the resummation of the amplitude, the final amplitude is then analytically continued to a region where the source is strong.

Without loss of generality, we consider only $pp$ in the rest of the paper as we expect the result at high energy is the same as that for $\pbar p$.  We first solve the classical field of each proton to the leading order in coupling constant and the strength of the source assuming the source is static.  Working in the covariant gauge, for a single proton, the equation of motion become Abelian \cite{Kovchegov1996} so that the total field of the two colliding protons is the superposition of the solution of a signle proton.  Solving the exact solution for two colliding hadrons is a much more difficult problem and the approximate solution to the next order has been found \cite{Kovner1995a,Kovner1995b,Kovchegov1997,Gyulassy1997}.  Nevertheless, a significant modification to the scattering amplitude can be found even in the leading order.  

With the inclusion of the classical effective theory, there are two types of gauge fields in the QCD Lagrangian: the classical field and the quantum gluons which are involved in the 2-to-2 scattering.  The cross term between the classical field and the quantum gluon in the Lagrangian introduces new interaction vertices in the Feynman diagram.  For the quantum gluon, we choose to work in the background gauge which is consistent with the covariant gauge for the classical field, as shown in Section \ref{sec:4}.  The classical field modifies only the gluon propagator.  The modified propagator to the leading order of the classical field is derived and shown to satisfy the gauge invariance condition.  We found that the gluon-gluon scattering amplitude is suppressed significantly in a strong background field; therefore, when the collinear gluon has a small $x$ value, the minijet cross section is highly suppressed.  This suppression is consistent with the possible breakdown of the collinear factorization formula in a high gluon density region.  In the small-$x$ region, the cross section should be calculated with un-integrated gluon density in the $k_\perp$ factorization scheme, instead of the parton model.  The modified amplitude provides the right amount of taming of the rise of minijet cross section and the total cross section can qualitatively describe the total cross section over the entire range of the available data, i.e. from $\sqrt{s}=5$ GeV to 30 TeV.  A detail analysis of the phenomenological implication will be presented in our recent proposed classical color field modified minijet model \cite{Cheung2011b}

The focus of this paper is to derive the gluon-gluon scattering correction due to the presence of the classical field of the colliding protons within the Gaussian approximation of CGC.  The remainder of the paper is organized as follows. In Sec. \ref{sec:2}, we derive the solution of the Yang-Mills equation of motion for the classical field of a proton.  We discuss how the observable is obtained using the IIM Gaussian approximation in Sec. \ref{sec:3}.  In Sec. \ref{sec:4}, the classical field correction to the gluon-gluon scattering amplitude is derived and shows that there is only the modification of the gluon propagator.  This work is concluded in Sec. \ref{sec:5}.  Appendix \ref{sec:GI} is devoted to discussions of gauge invariance in the present approach.  

\section{\label{sec:2} Classical field of a proton}
In this section, we discuss the classical field given by a specific source distribution $\rho_a(x)$ of a proton travelling near the speed of light.  Consider a proton moving along the positive light-cone ($+z$ direction) with velocity $v^\mu = (1,0,0,1)$ as a color source with current density $\rho$.  The color charge distribution inside the proton along the longitudinal direction is Lorentz contracted; therefore, it is localized near $x^- = 0$\footnote{Light-cone coordinate is used: $x_{LC}^\pm = \frac{1}{\sqrt{2}}(x^0\pm x^3)$ and $x_{LC}^i = x^i$.  The subscribe $LC$ will be omitted in our notation.}.  Due to time dilation, the source with larger $x$ appears to be static as far as the dynamics of the small $x$ gluon is concerned.  Thus, the source is independent of light-cone time $x^+$.  The source can be written as 
\begin{equation}
J^{a\,\mu} (x^-,x_\perp) = \delta^{\mu +}\rho^a(x^-,x_\perp).
\end{equation}
The classical field $A^{a\,\mu}$ is governed by the Yang-Mills equation of motion
\begin{equation}
D^{ab}_\nu F^{b\,\mu\nu} = J^{a\,\mu}. \label{eq:2:YM}
\end{equation}
Despite the non-linearity of eq.(\ref{eq:2:YM}), a close solution can be obtained with the covariant gauge $\partial_\mu A^\mu =0 $ and the static assumption.  
For a static source, $\partial_+ J^+ = 0$, it is consistent to look for a solution of $A$ that satisfies 
\begin{equation}
A_+ = A^- = 0. \label{eq:2:2}
\end{equation} 
The static condition also applies to $A$ so that $A$ is independent of $x^+$, $A^\mu=A^\mu(x^-,x_\perp)$; therefore, the partial derivative of $A$ with respect to $x^+$ vanishes, 
\begin{equation}
\partial_+ A^\mu = \partial^- A^\mu = 0. \label{eq:2:3}
\end{equation}
With eq. (\ref{eq:2:2}) and (\ref{eq:2:3}), the gauge condition together reduces to 
\begin{eqnarray}
\partial_+ A^+ +\partial_- A^- &-& \partial_i A_i =0 \nonumber\\
\Rightarrow  \partial_i A_i &=&0. \label{eq:2:4}
\end{eqnarray}
Due to the finite size of the proton, we use the boundary condition of the field at infinity that $A(|x_\perp| \rightarrow \infty) =0 $.  Thus, we have
\begin{equation}
A_i(x^-,x_\perp) = 0 .
\end{equation}
Only $A^+=A_-$ is non-zero which also implies $A^a_\mu A^{b\,\mu} = A^{a\,-} A^{b\,+}+A^{a\,+} A^{b\,-} - A^{a\,i} A^{b\,i} = 0$.  These conditions greatly simplify eq. (\ref{eq:2:YM}).  Eq. (\ref{eq:2:YM}) for $\mu = +$ reduces to 2D Possion equation,
\begin{equation}
\nabla_\perp^2 A^{a\, +}(x^-,x_\perp) = \rho(x^-,x_\perp) ,\label{eq:2:5}
\end{equation}
and the solution is 
\begin{equation}
A_1^{a\, +}(x^-,x_\perp) = -\int \frac{d^2y_\perp  }{(2\pi)^2}d^2k_\perp \frac{\rho_1^a(x^-,y_\perp)}{k_\perp^2}e^{ik_\perp \cdot (x-y)_\perp}.\label{eq:2:A1}
\end{equation}
We assign an index 1 to $\rho$ to indicate that the source is moving toward $+z$.  For a source $\rho_2$ moving toward the opposite direction, the static assumption results $A^+=0$ and $\partial^+ A_\mu =0$.  The only non-zero field is 
\begin{equation}
A_2^{a\, -}(x^+,x_\perp) = -\int\frac{d^2y_\perp}{(2\pi)^2}  d^2k_\perp \frac{\rho_2^a(x^+,y_\perp)}{k_\perp^2}e^{ik_\perp \cdot (x-y)_\perp}. \label{eq:2:A2}
\end{equation}
In general, if there are two approaching sources, there will be both non-zero $A^+$ and $A^-$.  Due to color precession of the sources, the sources cannot be treated as static anymore.  The field of each source will induce a change in the other source.  Since we are working toward the first order correction of a bare vacuum due to the source, these higher order cross inductions will be ignored.  Therefore, the total field to the leading order approximation of the source $\delta^{\nu+}\rho_1^{a\, +} + \delta^{\nu -}\rho_2^{a\, -}$ is 
\begin{equation}
A^{a\,\mu} = \delta^{\mu +}A_1^{a\,+}[\rho_1] + \delta^{\mu -}A_2^{a\,-}[\rho_2] + O(\rho_1\rho_2) +\cdots
\end{equation}

\section{\label{sec:3} Gaussian Average}
As in the CGC approach, our source $\rho$ is treated as a random variable.  A physical observable $O$ is calculated by first obtaining $O = O[\rho]$ in terms of $\rho$, then averaging over $\rho$ with a weight functional $W[\rho]$.  Using the IIM Gaussian Ansatz \cite{Iancu2002b}, the observable $O$ is given by
\begin{equation}
\left\langle O\right\rangle =  N\int \mathcal{D}\rho\, O[\rho] \, W[\rho], \label{eq:3:CGC}
\end{equation}
where
\begin{equation}
W[\rho] = \exp\left\{ -\int dy d z_{1_\perp} d z_{2_\perp} \frac{\rho^a_y(z_{1_\perp})\rho^a_{y}(z_{2_\perp})} {2\lambda_y(z_{1_\perp},z_{2_\perp})} \right\} \label{eq:3:W0},
\end{equation}
$N$ is normalization constant and $\lambda_y(z_{1_\perp},z_{2_\perp})$ characterizes the correlation between two positions in the source.

The source will transform if we switch gauge, $\rho\rightarrow U \rho U^\dagger$.  The Gaussian weight function is gauge invariant and $\mathcal{D}\rho$ is an invariant measure.  Therefore, the observable is gauge invariant if the observable $O[\rho]$ is also invariant when it is evaluated with a gauge transformed source $\rho'=U \rho U^\dagger$, therefore, $O[\rho] =O[\rho']$.  We will use this criterion to check gauge invariance of the scattering amplitude in the classical field.  The detail is presented in Appendix \ref{sec:GI}.

The weight function in IIM model \cite{Iancu2002b} is expressed in terms of variable $y=\ln (z^-/z_0)$ and $z_\perp$.  For our purpose, it is more convenient to use coordinate variables $z = \{z^-, z_\perp\}$ as 
\begin{equation}
W[\rho] = \exp\left\{ -\int dz^- d z_{1_\perp} d z_{2_\perp} \frac{\rho^a(z^-,z_{1_\perp})\rho^a(z^-,z_{2_\perp})} {2\lambda'(z^-,z_{1_\perp},z_{2_\perp})} \right\} \label{eq:3:W},
\end{equation}
where 
\begin{equation}
\lambda'(z^-) = \frac{\lambda_y}{z^-}.
\end{equation}
 The function $\lambda(z^-)$ has a support for $0<z_1^-<L^- \sim 1/(xP^+)$ where $xP^+$ is the momentum of the collinear gluon involved in the scattering.  The $x$ value can be regarded as the scale that separates the source with $x_{source} > x$ and the field with $x_{field}<x$.  A corresponding correlation function in the transverse plane, $\mu$, can be defined as the integral of $\lambda$ over $y$, 
\begin{equation}
\mu_\tau(z_{1_\perp}-z_{2_\perp})=\int_{-\infty}^{\tau(x)} dy \lambda_y(z_{1_\perp}-z_{2_\perp}),
\end{equation}
and its derivative relates to $\lambda(z^-)$ as
\begin{eqnarray}
\lambda_y = \frac{\partial \mu_y}{\partial y} \, \Rightarrow \,\frac{\partial \mu(z^-)}{\partial z^-} = \frac{\lambda_y}{z^-} =  \lambda'(z^-).
\end{eqnarray}
The two-point correlation function in space-time coordinates becomes
\begin{equation}
\langle \rho^a(z_1) \rho^b(z_2) \rangle= \delta^{ab}\delta(z_1^--z_2^-) \frac{\partial \mu}{\partial z^-}(z^-,z_{1_\perp}-z_{2_\perp})\label{eq:3a:rho}, 
\end{equation}
Since there is a one-to-one correspondence between the coordinate $z^-$ and rapidity $\tau = \ln(1/x)$ and, in turn, $x$, we translate the longitudinal dependence into the rapidity dependence, namely, 
\begin{equation}
\frac{\partial \mu}{\partial z^-}(z^-,z_{1\perp}-z_{2\perp})= \frac{\partial \mu_x}{\partial z^-}(z_{1\perp}-z_{2\perp})
\end{equation}
so that $\mu_x = \int_0^{L^-} \frac{\partial \mu_x}{\partial z^-} = \frac{\partial \mu_x}{\partial z^-}\, L^-$
where $L^- \sim 1/(x P^+)$ is the longitudinal upper limit of the source.
Simplifying the notation of the two-point correlation, we have
\begin{equation}
\langle \rho^a(z_1) \rho^b(z_2) \rangle= \delta^{ab}\delta(z_1^--z_2^-)\bar{\mu}_x(z_{1_\perp}-z_{2_\perp}) \label{eq:3:rho},  
\end{equation}
where $\bar{\mu}_x \equiv \mu_x/L^-$.  If we integrate this correlation over the longitudinal directions $z_1^-$ and $z_2^-$, we recover the same correlation in eq. (3.5) of \cite{Iancu2002b}. 

The two-point correlation also provides that the Gaussian average of any term with odd power of $\rho$ is zero but terms with even power of $\rho$ is non-zero.  In our calculation, since the field $A$ is linear to $\rho$ (see eq.(\ref{eq:2:A1})), any non-zero leading order contribution must come from $A_1^2$ or $A_2^2$ because the sources from two different protons do not correlate; therefore, $\langle A_1[\rho_1]A_2[\rho_2] \rangle =0$.  In terms of the Feynman diagram, the leading order correction corresponds to the diagram with two interaction vertices with the classical field.  

\section{\label{sec:4} Interaction between classical and quantum gluons}
With the presence of the classical field due to the proton, we write the total gluon field as the sum of the classical field $A$ and the quantum field $B$.  $A$ is treated as a prescribed field, a background, while $B$ is the quantum gluon which is involved in scattering processes.  The interactions between $A$ and $B$ are naturally emerged from the QCD Lagrangian.  

\subsection{Lagrangian}
We write the full Lagrangian, ignoring the fermion field, as the classical QCD Lagrangian with $A$ replaced by $A+B$
\begin{equation}
 \mathcal{L}_{QCD} = -\frac{1}{4} F^{a\,\mu\nu}F_{\mu\nu}^{a} +  J^{a\,\mu}(A+B)^{a}_{\mu},
\label{eq:4:L0}
\end{equation}
where
\begin{eqnarray}
 F^{a\,\mu\nu} = &&\partial^\mu (A+B)^{a\,\nu} -\partial^\nu (A+B)^{a\,\mu}\nonumber \\
&& +gf^{abc} (A+B)^{b\,\mu} (A+B)^{c\,\nu},
\end{eqnarray}
$J$ is the classical source of the two protons.  The Lagrangian is invariant under the infinitesimal gauge transformation, 
\begin{equation}
\left\{ \begin{array}{lll}
J^{a\,\mu} &\rightarrow & J'^{a\,\mu} = (\delta^{ab}-f^{abc}\alpha^c) J^{b\,\mu}\\
(A+B)^a_\mu &\rightarrow &(A'+B')^a_\mu =(A+B)^a_\mu \\
&&+ f^{abc}(A+B)^b_\mu \alpha^c +\frac{1}{g}\partial^\mu \alpha^a
\end{array} \right. \label{eq:4:GT1}
\end{equation}
provided that $\partial_\mu J ^\mu = 0$ which is satisfied by a static source.  A detailed discussion of the gauge invariance of the theory will be presented in Section \ref{sec:GI}.  

We can organize the first terms of the Lagrangian in eq. (\ref{eq:4:L0}) in terms of the powers of $B$,
\begin{equation}
  \mathcal{L}_{gauge} \equiv -\frac{1}{4} F^{a\,\mu\nu}F_{\mu\nu}^{a} = \mathcal{L}_0+\mathcal{L}_1+\mathcal{L}_2+\mathcal{L}_3+\mathcal{L}_4
\end{equation}
where the subscripts represent the order of $B$ contained in each term.  $\mathcal{L}_0$ is independent of $B$.  Upon volume integration, it becomes a constant in the action, so it does not affect the observables and can be ignored.  The first order term is 
\begin{eqnarray}
\mathcal{L}_1 &=& - \bar{D}^{ab}_\nu \,\bar{F}^{b\, \mu \nu} B_\mu^a + \mbox{(total derivative)} \nonumber\\
 &=& -J^{a\, \mu} B^a_\mu + \mbox{(total derivative)},
\end{eqnarray}
according to the field equation of the classical field, eq. (\ref{eq:2:YM}), where $\bar{D}$ is the covariant derivative involving only $A$ 
\begin{equation}
 \bar{D}^{ac}_\mu  = \left(\partial_\mu \delta^{ac} + gf^{abc}A^{b}_\mu\right)
\end{equation}
and $\bar{F}$ is the field tensor of $A$,  
\begin{align}
\bar{F}^{a\, \nu \mu}& = \partial^\mu A^{a\,\nu} -\partial^\nu A^{a\,\mu} +gf^{abc} A^{b\,\mu} A^{c\,\nu}.
\end{align}
$\mathcal{L}_1$ cancels with $J\cdot B$ in the Lagrangian so that there is no single leg vertex diagram for $B$.  $\mathcal{L}_3$ and $\mathcal{L}_4$ contain terms proportional to \[gB^3, \quad g^2 AB^3, \quad g^2B^4.\] The first type is the original QCD three-gluon vertex.  The last two types are higher order terms in the coupling $g$.  Our interest, instead, is in the leading order classical field effect which comes from the quadratic term as
\begin{equation}
  \mathcal{L}_2 = \frac{1}{2} B^a_\mu \left[ g^{\mu\nu}\bar{D}_{ac}^\rho \bar{D}_{cb\,\rho} -\bar{D}_{ac}^\mu \bar{D}_{cb}^{\nu} - 2 g f^{acb}\bar{F}^{c\,\mu\nu}\right] B_\nu^b.
\end{equation}
\subsection{Background gauge}
It has been shown that if one introduces a classical gauge field to the QCD Lagrangian, it is convenient to fix the gauge with background gauge \cite{Abbott1981}.  We choose the gauge fixing function to be 
\begin{equation}
f^a \equiv \bar{D}^{ab}_\mu (A+B)^{b\,\mu} =  \bar{D}^{ab}_\mu B^{b\,\mu}  \label{eq:4:2},
\end{equation}
where the last simplification is valid because $A$ satisfies $\partial _\mu A^\mu =0$ and $A_\mu A^\mu = 0$; thus,  the classical field also satisfies background gauage condition,
$\bar{D}^{ab}_\mu A^{b\,\mu} = \partial_\mu (A_1^{a\,\mu}+A_2^{a\,\mu}) +gf^{abc}(A_1+A_2)^b_\mu(A_1+A_2)^{c\,\mu} =g f^{abc}(A_1^{b\,+}A_2^{c\,-}+A_1^{c\,+}A_2^{b\,-}) =0$.

\subsection{Quadratic term}
Including the background gauge fixing term 
\begin{eqnarray}
  \mathcal{L}_{GF} =&& -\frac{1}{2} (\bar{D}_{ab}^\mu B^b_\mu)^2 \nonumber \\
=&& \frac{1}{2} B^a_\mu \bar{D}_{ac}^\mu \bar{D}_{cb}^{\nu} B^b_\nu 
+ \mbox{total derivative},
\end{eqnarray}
the quadratic term in $B$ of the Lagrangian becomes
\begin{align}
  \mathcal{L'}_2=&\mathcal{L}_2+\mathcal{L}_{GF} \nonumber \\
 =& \frac{1}{2} B^a_\mu \left[ g^{\mu\nu}\bar{D}_{ac}^\rho \bar{D}_{cb\,\rho} - 2 g f^{acb}\bar{F}^{c\,\mu\nu}\right] B_\nu^b \nonumber \\
=& \frac{1}{2} B^a_\mu \left[ g^{\mu\nu} \left(  \delta^{ab}\Box + gf^{acb} 2A^c_\rho \partial^\rho \right.\right. \nonumber \\
&+ \left. \left. g^2 f^{aec}f^{cdb}A^e_\rho A^{d\,\rho} \right) - 2 g f^{acb}\bar{F}^{c\,\mu\nu} \right] B_\nu^b . \label{eq:4:L2}
\end{align}
The last three terms in eq. (\ref{eq:4:L2}) are the interactions of interest.  The term $ g g^{\mu\nu} f^{acb} B^a_\mu A^c_\rho \partial^\rho B^b_\nu$ is the vertex with one classical leg and two quantum legs.  Since $A= A_1+A_2$, this term actually represents two types of interactions, one with each proton.  An interaction with only one power of $A$ does not directly contribute because of the vanishing Gaussian average $\langle A \rangle \propto \langle \rho \rangle =0$.  Therefore, we will expect the gluon to interact with the same classical field through this interaction twice to give a first non-zero correction at the $g^2 A^2$.  The next term is a vertex with four legs while two of them are classical.  This term involves $A\cdot A$.  Since $A_1\cdot A_1 = A_2\cdot A_2 = 0$, it leaves with $g^2 A_1 \cdot A_2$.  $A_1$ and $A_2$ belongs to two different Gaussian averaging procedures, $\langle A_1 A_2 \rangle = \langle A_1 \rangle \langle A_2 \rangle=0$.  So we need to have one more interaction with either classical field.  However, this term is already at $g^2 A^2$ order.  Any extra interaction will give one order higher correction.  It will be ignored in our first order calculation.  For the last term, we can write the field strength into the sum of field strength due to proton 1, $F_1$, proton 2, $F_2$ and the cross term $F_{12}$.  Explicitly, the cross term is 
\[gF_{12}^{c\,\mu\nu}=g^2f^{cde}(A_1^{d\,\mu}A_2^{e\,\nu}+A_2^{d\,\mu}A_1^{e\,\nu}).\]
It also has a vanishing leading order contribution.  The important consequence of not having cross term in the leading order is that the contributions of proton 1 and 2 can be calculated independently.

From here on, we will only consider the proton moving to $+z$ direction.  The following results can be easily converted to the case with the proton moving to $-z$ direction.  The classical field $A$ has only "+" component and the derivative w.r.t. $x^+$ vanishes $\partial_+A^+ = \partial^-A^+ = 0$, so $A^a_\mu A^{b\,\mu} = 0$ and only the $+i$ and $i+$ components, $\bar{F}^{a\,i+}=-\bar{F}^{a\,+i}=\partial^i A^{a\,+}$, do not vanish.  Eq. (\ref{eq:4:L2}) is simplified to 
\begin{align}
  \mathcal{L'}_2=& \frac{1}{2} B^a_\mu \left[ g^{\mu\nu} (\delta^{ab}\square - gf^{abc} 2A^{c\,+} \partial^- ) \right. \nonumber \\
&+ \left. 2 g f^{abc} \sum_{i=1}^2(g^{\mu
-}g^{\nu i}-g^{\mu i}g^{\nu -})\partial_i A^{c\,+}  \right] B_\nu^b . \label{eq:4:L2b}
\end{align}
The inverse of $g^{\mu\nu} \delta^{ab}\square$ in the first term gives the Feynman propagator.  Let us denote the second term and the third term as 
\begin{align}
\mathcal{L}_{int_1} = &-g\, g^{\mu\nu}f^{abc} B^a_\mu A^{+c}\partial^- B^b_\nu \label{eq:4:Lint1}\\
\mathcal{L}_{int_2} =& gf^{abc}\sum_{i=1}^2 B^a_\mu  (g^{\mu -}g^{\nu i}-g^{\mu i}g^{\nu -})\partial_i A^{c\,+} B^b_\nu \label{eq:4:Lint2}\\
\mathcal{L}_{int} =&\mathcal{L}_{int_1} +\mathcal{L}_{int_2} 
\end{align} 
There are two kinds of interactions: (1) the gluon changes color but not polarization and (2) the gluon
changes both color and polarization.  The Feynman rule of these classical-quantum vertices are for vertex (1): 
\begin{equation}
g g^{\mu\nu} f^{abc}  q^- A^{+c}(p-q),
\end{equation}
and for vertex (2): 
\begin{equation}
-gf^{abc}\sum_{i=1}^2 (g^{\mu -}g^{\nu i}-g^{\mu i}g^{\nu -}) (p_i-q_i) A^{c\,+}(p-q),
\end{equation}
where the incoming and outgoing gluons are labeled by ${p,\mu,a}$ and ${q,\nu,b}$, respectively.  $A^{+c}(p-q)$ is the Fourier transform of the solution of $A$ obtained in section \ref{sec:2}.  Since $A^{+c}(x^-,x_\perp)$ is independent of $x^+$, $A^{+c}(p-q) \propto \delta(p^--q^-)$.  This delta function turns out to be very crucial to simplify the calculation as discussed below.  

\subsection{Vertex and propagator corrections}
\begin{figure}[h]
\includegraphics[width=0.6\linewidth]{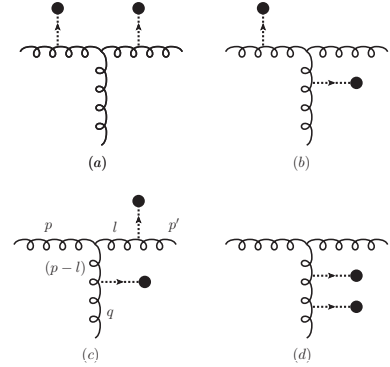}
\caption{(a), (b) and (c) are the vertex corrections and (d) is the propagator correction due to the classical-quantum interaction.  A dotted line ending with a black dot represents the interaction with $A^{+}$.}
\label{fig:4:1}
\end{figure}

In pQCD, the high energy gluon-gluon elastic scattering cross section is dominated by the $t$- and $u$-channel gluon exchanges.   
For these exchange, the classical-quantum interaction introduces vertex and propagator corrections to the Feynman diagrams as illustrated in Fig. \ref{fig:4:1}.
\begin{figure}[h]
\includegraphics[width=0.7\linewidth]{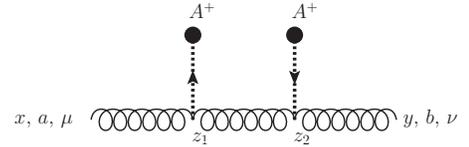}
\caption{Feynman diagram of the propagator in classical field.  Each vertex can be $\mathcal{L}_{int_1}$ or $\mathcal{L}_{int_2}$.  The dotted line represents the momentum exchange between the gluon and the classical field.}
\label{fig:4:2}
\end{figure}
Thanks to the delta function, $\delta(k^-)$ of $A^{+}(k)$, (a), (b) and (c) do not contribute to the S-matrix.  To see this, consider diagram \ref{fig:4:1}(c).  The amplitude is proportional to 
\begin{eqnarray}
\delta(l^-&& -p'^-)\delta(p^--l^--q^-)\frac{1}{l^2+i\epsilon} \frac{1}{(p-l)^2+i\epsilon} dl^+ \nonumber \\
\propto&& \frac{1}{l^+ +\frac{-l_\perp^2+i\epsilon}{2l^-}}\, \frac{1}{l^+ -p^+ +\frac{-(l-p)_\perp^2+i\epsilon}{2(l^--p^-)}} dl^+.\label{eq:4:3}
\end{eqnarray}
$p'$ is on-shell so $l^- = p'^->0$.  As we will see later,  after Gaussian averaging, the two classical lines are in fact connected with a momentum $k^+$ and $k_\perp$ going through.  The momentum $p$, $p'$ and $q$ satisfy momentum conservation.  Therefore, $(p-q)^2=p'^2$ and it implies that $q^- = q^2/2p^+ < 0$ as $q^2 <0$ in the physical region.  We also have $l^--p^- = -q^- >0$.  So both poles in eq. (\ref{eq:4:3}) are in the lower half plane, provided that there is not any $l^+$ in the numerator.  By choosing the contour to close on the upper half plane, one obtains zero contribution.  This mathematical property applies also in diagram (b) and (c).  Similar argument has been made in an earlier study on photon production of quark \cite{Gelis2002a}.  
Therefore, there is no vertex correction due to the classical field.  

For diagram (d), the momentum integration of the loop in the propagator goes like $\int dl^+/(l^+ - l_\perp^2/2l^- + i\epsilon/2l^-)$ so it does not vanish, independent of the choice of the contour.  

To calculate the propagator correction, we write down the two-point Green's function to the leading order in $A[\rho]$ as 
\begin{align}
\langle G_{\mu\nu}^{ab}&(x,y;\rho)\rangle 
=\langle T\{ B^a_\mu(x) B^b_\nu(y)\} \rangle \nonumber\\
&+\langle T\{ B^a_\mu(x) \frac{i^2}{2!}\int
d^4{z_1}d^4{z_2}\mathcal{L}_{int}(z_1) \mathcal{L}_{int}(z_2)
 B^b_\nu(y)\} \rangle
\end{align}  
where the notation $\langle \cdots \rangle$ means
Gaussian averaging over the classical source defined in eq. (\ref{eq:3:CGC}).  The interaction term linear in
$A^+$ vanishes because $\langle A^+(x) \rangle = 0$.  The first term is the Feynman propagator in bare vacuum.  The second term represents Feynman diagrams as in Fig. \ref{fig:4:2}.

There are totally four different contributions coming from the binary pair of interactions $\mathcal{L}_{int_1}$ and $\mathcal{L}_{int_2}$.  We will call them vertex 1 and 2, respectively.  and denote the propagator correction involving interactions with two $\mathcal{L}_{int_1}$ as 1-1 term, with two $\mathcal{L}_{int_2}$ as 2-2 term and with the mixed vertices as 1-2 and 2-1 terms.  

\subsection{$\langle A^{a\,+}A^{b\,+} \rangle$}
A crucial ingredient of the calculation is the random source averaging of $A$.  Applying the IIM model, eq. (\ref{eq:3:rho}), on $A^{a\,+}A^{b\,+}$ of eq. (\ref{eq:2:A1}) gives
\begin{eqnarray}
&&\langle A^{a\,+}(z_1)A^{b\,+}(z_2) \rangle \nonumber \\
&&= \frac{\delta^{ab}}{(2\pi)^2} \delta(z_1^- - z_2^-) \int \frac{d^2k_\perp}{k_\perp^4} \bar{\mu}_x(k_\perp^2) e^{ik_\perp \cdot (z_1-z_2)_\perp}\label{eq:4:AA},
\end{eqnarray}
where $\bar{\mu}_x(k_\perp^2)$ is the Fourier transform of $\bar{\mu}_x(x_\perp-y_\perp)$.

\subsection{1-1 term}
Let us demonstrate the calculation of the propagator due to gluon interacting with $\mathcal{L}_{int_1}$ twice at two positions.  Let $p_1$, $p_2$ and $p_3$ be the momentum of the gluon lines on the left, middle and right, respectively, in Fig. \ref{fig:4:2}.  The Two-point function is given by
\begin{widetext}
\begin{align*}
G^+_{11}&\,^{ab}_{\mu\nu}(x,y) 
= 2\times\frac{i^2}{2!}\langle T\{\int d^4z_1 d^4z_2 B^a_\mu(x) \mathcal{L}_{int_1}(z_1) \mathcal{L}_{int_1}(z_2) B^b_\nu(y)\}\rangle\\
=&-\langle T\{\int d^4z_1 d^4z_2 B^a_\mu(x) (-g f^{cmd} B^c_\rho(z_1) A^{m\,+}(z_1) \partial^- B^{d\,\rho}(z_1))
(-g f^{enf} B^e_\sigma(z_2) A^{n\,+}(z_2) \partial^- B^{f\,\sigma}(z_2)) B^b_\nu(y)\}\rangle
\end{align*}
\end{widetext}
There are four different ways to contract the fields.  The first way is to contract \[
B^a_\mu(x) B^c_\rho(z_1), \quad B^{d\,\rho}(z_1)B^e_\sigma(z_2),\quad B^{f\,\sigma}(z_2) B^b_\nu(y),\] which gives
\begin{align}
G^{+}_{11}&^{(1)}\,^{ab}_{\mu\nu}(x,y) \nonumber \\
=& -g^2 g_{\mu\nu} f^{amd} f^{dnb} \int d^4z_1 d^4z_2 \langle A^{m\,+}(z_1)A^{n\,+}(z_2) \rangle\nonumber \\
&\times \frac{d^4 p_1}{(2\pi)^4}\frac{d^4 p_2}{(2\pi)^4}\frac{d^4 p_3}{(2\pi)^4} \frac{-i}{p_1^2} \frac{-i}{p_2^2} \frac{-i}{p_3^2} \nonumber \\
& \times e^{-ip_1\cdot(x-z_1)} (-ip_2^-) e^{-ip_2\cdot(z_1-z_2)}  (-ip_3^-) e^{-ip_3\cdot(z_2-y)}
\end{align}
The Gaussian average of $A^mA^n$ produces $\delta^{mn}$ so that $f^{amd} f^{dnb}$ becomes $f^{amd} f^{dmb} = -N_c \delta^{ab}$.  In order to carry out the integration of $z_1$ and $z_2$, we write $\langle AA \rangle$ in 4D fourier transform as 
\begin{eqnarray}
\langle &&A^{a\,+}(z_1)A^{b\,+}(z_2) \rangle \nonumber\\
&& = \delta^{ab} \int d^4k_1 d^4k_2 e^{-ik_1\cdot z_1} e^{-ik_2\cdot z_2} F(k_1,k_2), \label{eq:4:AA2}
\end{eqnarray}
where 
\begin{eqnarray}
&&F(k_1,k_2) \nonumber \\
&&= \frac{\bar{\mu}_x(k_{1\perp}^2)}{(2\pi)^3} \frac{\delta(k_1^-)\delta(k_2^-)\delta^2(k_{1\,\perp} + k_{2\,\perp}) \delta(k_{1}^+ + k_{2}^+)} {k_{1\,\perp}^2 k_{2\,\perp}^2}.
\end{eqnarray}
With some algebraic manipulations, 
\begin{align}
G^{+\,(1)}_{11}\,^{ab}_{\mu\nu}(x,y) = & -ig^2g_{\mu\nu}\delta^{ab} N_c  (p_2^-)^2  
\frac{d^4 p_1 d^4 p_2 d^4 p_3}{(2\pi)^4}\nonumber \\
& \frac{1}{p_1^2 p_2^2 p_3^2} e^{-ip_1\cdot x + ip_3\cdot y} F(p_1-p_2, p_2-p_3).
\end{align}
$p_2^-$ and $p_3^-$ combine to be $(p_2^-)^2$ because of the $\delta(k^-) =\delta(p_2^--p_3^-)$ in $F$ which implies that the classical field does not carry momentum at the $-$ component.  Thus,
\begin{equation}
p_1^-=p_2^-=p_3^-.  \label{eq:4:p123}
\end{equation} 
Integrating over $p_2$, we have
\begin{align}
&\int d^4p_2 \frac{1}{p_2^2+i\epsilon} F(p_1-p_2, p_2-p_3) \nonumber \\
&= \frac{\delta^4(p_1-p_3)}{(2\pi)^3}  \int \frac{dp_2^+}{2p_2^+ p_1^- -p_{2\perp}^2 +i\epsilon} \int \frac{d^2 k_\perp }{k_\perp^4}\bar{\mu}_x(k_\perp^2),
\end{align}
where we have expanded $p_2^2$ in the denominator, integrated $p_2^+$ with the delta function $\delta(p_1^--p_2^-)$ and replaced $p_{2\perp} - p_{1\perp}$ with $k_\perp$.  The $p_2^+$ integration can be evaluated by closing the contour on either the upper or the lower half plane (one can also calculate the principal value of the integration from $-L$ to $L$ as $L\rightarrow \infty$),  
\begin{align}
\int \frac{dp_2^+}{2p_2^+ p_1^- -p_{2\perp}^2 +i\epsilon} = -i\pi\frac{\theta(p_1^-) - \theta(-p_1^-)}{2p_1^-}. \label{eq:4:dp+}
\end{align}
The origin of the $k_\perp$ integration is from the Fourier transform of the transverse distribution of the classical source $\rho$ (see eq. (\ref{eq:2:A1})).  When a gluon with momentum $Q$ probes the transverse distribution of the source, it defines the smallest size that can be resolved.  Therefore, the spatial integration of the source is from $1/Q$ to the radius of a proton, $R_p$.  In momentum space, $Q^2$ porives a UV-cutoff of the integration of the $k_\perp^2$.  The IR-cutoff is set to be $1/R_p^2$.  Therefore, it becomes 
\begin{align}
I(Q^2,x)=\int_{1/R_p^2}^{Q^2} \frac{\bar{\mu}_x(k^2_\perp)}{k_\perp^4} d^2 k_\perp, \label{eq:4:dk}
\end{align}
where the explicit form of $I(Q^2,x)$ depends on the $k^2_\perp$ dependence of $\bar{\mu}_x$.  
The the two-point function is  
\begin{align}
G^{+\,(1)}_{11}\,^{ab}_{\mu\nu}(x,y) = &-\frac{\alpha_s N_c}{4\pi } I(Q^2,x) (\theta(q^-)-\theta(-q^-))\nonumber \\
& g_{\mu\nu}\delta^{ab} \frac{d^4 q}{(2\pi)^4} \frac{q^-}{q^4} e^{-i q\cdot (x -y)}.
\end{align}
Now we consider the other way to contract the $B$ fields.  As we mentioned above, eq. (\ref{eq:4:p123}), that all the $-$ components of the momentum are the same.  This makes the results of contracting the $B$ fields in the different ways identical.  For example, if one contracts $B^a_\mu(x) B^{d\,\rho}(z_1)$ and $B^c_\rho(z_1) B^e_\sigma(z_2)$.  The color index of $f^{cmd}$ becomes $f^{dmc}$ so it picks up a negative sign.  However, the $\partial^-$ is now acting on the gluon line with $p_1$ at $z_1$, $\partial^-e^{-ip_1(x-z_1)} = +ip_1^-$.  It is different by another negative sign.  So there is no over all sign change between the two different contraction.  As $p_1^- = p_2^-$, the two contraction become exactly the same.  So the contribution due to the 1-1 term is 4 times of the result from any one of the contractions.  The propagator correction in momentum space is given by
\begin{equation}
G^+_{11}\,^{ab}_{\mu\nu}(q)=-g_{\mu\nu}\delta^{ab}\frac{\alpha_s N_c}{\pi} I(Q^2,x)(\theta(q^-)-\theta(-q^-)) \frac{q^-}{q^4}.
 \label{eq:4:G11}
\end{equation}

\subsection{1-2 and 2-1 terms} 
As we have already worked out the 1-1 term, the details for the calculations of the rest of the terms are similar.  We will point out a few keys in the calculation.  For the 1-2 and 2-1 terms, The interaction term $\mathcal{L}_{int_2}$ consists of a $\partial_i$ on the classical field.  In the 1-2 term, $\mathcal{L}_{int_1}$ is at $z_1$ and $\mathcal{L}_{int_2}$ is at $z_2$.  The derivative in $\mathcal{L}_{int_2}(z_2)$ acts at position $z_2$ of $\langle A(z_1)A(z_2)\rangle$ and gives $-ik_\perp$.  While the 2-1 term with the same order of contraction will have the derivative acting on $z_1$ resulting $+ik_\perp$.  Each contraction in the 1-2 term will cancel with that in the 2-1 term. 

\subsection{2-2 term}
The interaction vertex is symmetric under exchange of color index $a\leftrightarrow b$ together with Lorentz index $\mu \leftrightarrow \nu$.  So there are totally 8 different contractions, including exchanging $z_1$ and $z_2$, that give identical contributions.  At the level of the Lorentz index, the only surviving term is $\sum_{i,j}-g^{ij}g_\mu^-g_\nu^-$.  The 2-2 term is 
\begin{align}
G^+_{22}\,^{ab}_{\mu\nu}(x,y) 
=& \frac{4g^2N_c \pi}{(2\pi)^7}g_{\mu}^-g_{\nu}^-\delta^{ab}
\int\frac{d^4 p_1}{2p_1^- p_1^4} e^{-ip_1\cdot (x- y)} \nonumber \\ 
&\times (\theta(p_1^-) -\theta(-p_1^-))\int \frac{\bar{\mu}_x(k^2_\perp)}{k_\perp^2}d^2 k_\perp
\end{align}
Therefore, the propagator correction in momentum space is
\begin{equation}
G^+_{22}\,^{ab}_{\mu\nu}(q) = \frac{\alpha_s N_c }{\pi} g_{\mu}^-g_{\nu}^- \delta^{ab}
\frac{I'(Q^2,x)}{q^- q^4} (\theta(q^-) -\theta(-q^-)) \label{eq:4:G22},
\end{equation}
where $I'(Q^2,x) = \int \frac{\bar{\mu}_x(k^2_\perp)}{k_\perp^2} d^2 k_\perp$.  

\section{\label{sec:5}Result and discussion}
We have found the leading order correction of the gluon-gluon $t$ and $u$-channel scattering amplitude due to the classical color field in the background gauge only contributes to the propagator.  The classical field introduces two type of interactions to the gluon.  As for the propagator, single interaction with the classical field does not contribute because of the color neutrality assumption, namely the overall average of color field should be zero.  However, the fluctuation can be non-zero so that second order terms contribute.  Among all the second order diagrams, only the 1-1 and 2-2 terms survive.  The 1-1 term is diagonal in both color and Lorentz structure, while the 2-2 term contains $g_{\mu}^-g_{\nu}^-$.  When one tries to sum a series of 1-1 and 2-2 terms, any series with more than one 2-2 term will be zero.  It is because when one connect these diagram with a bare propagator, $g^{\nu \rho}$ is inserted between two diagrams.  Therefore, a 2-2 term connecting with a 2-2 term has a form of 
\begin{equation}
g_{\mu}^-g_{\nu}^- \times g^{\nu \rho} \times g_{\rho}^-g_{\lambda}^- = g_{\mu}^- g^{- -}g_{\lambda}^- = 0 \mbox{ since $g^{--}$ =0.}
\end{equation}
\begin{figure}[h]
\includegraphics[width=0.5\linewidth]{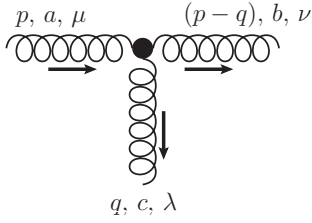}
\caption{Three gluon vertex.}
\label{fig:4:3}
\end{figure}
Furthermore, under the eikonal approximation at high energy, the three-gluon vertex shown in Fig. \ref{fig:4:3} becomes
\begin{equation}
2g\, g^{\mu\nu} p^\lambda f^{acb}.
\end{equation}
Contracting the 2-2 term to the gluon line with momentum $q$ forces $p^\lambda$ to be $p^-$ and the vertex vanishes because $p^-=0$ for an incoming gluon moving along the positive light-cone.  Therefore, when one applies the modified propagator to high energy gluon-gluon scattering, one could ignore the 2-2 term and keep only the 1-1 term.  

The difference between the protons moving at the $+z$ and the $-z$ direction is that for the one moving at the $-z$ direction, the $A$ field is $A^\mu=A^-$ instead of $A^+$.  To obtain the interactions, one just needs to exchange the index $+$ $\leftrightarrow$ $-$ in any fields and derivatives from the interaction term of the $+z$ case.  We denote the propagators of the case for $+z$ and $-z$ as $G^+$ and $G^-$, respectively.

The classical field modified gluon propagator to the first leading order to the field $A= A_1^++A_2^-$ is 
\begin{eqnarray}
G_{cl}\,^{ab}_{\mu\nu}(q) = &&G_{0}\,^{ab}_{\mu\nu}(q) + G^+_{11}\,^{ab}_{\mu\nu}(q) +G^-_{11}\,^{ab}_{\mu\nu}(q) \nonumber \\ 
&&+ G^+_{22}\,^{ab}_{\mu\nu}(q)+ G^-_{22}\,^{ab}_{\mu\nu}(q), \label{eq:4:Gcl}
\end{eqnarray}
where $G_{0}\,^{ab}_{\mu\nu}(q) = \frac{-i}{q^2}g_{\mu\nu}\delta^{ab}$ is the bare propagator in vacuum, 
\begin{equation}
G^\pm_{11}\,^{ab}_{\mu\nu}(q)=-g_{\mu\nu}\delta^{ab}\frac{\alpha_s N_c}{\pi} I(Q^2,x_{\pm})(\theta(q^\mp)-\theta(-q^\mp)) \frac{q^\mp}{q^4}.
 \label{eq:4:G11b}
\end{equation}
and,
\begin{equation}
G^\pm_{22}\,^{ab}_{\mu\nu}(q) = \frac{\alpha_s N_c }{\pi} g_{\mu}^\mp g_{\nu}^\mp \delta^{ab}
\frac{I'(Q^2,x_{\pm})}{q^\mp q^4} (\theta(q^\mp) -\theta(-q^\mp)) \label{eq:4:G22b},
\end{equation}
where $x_+$ and $x_-$ are the $x$ value of the source moving in the $+z$ and $-z$ direction, respectively.

So far, this is only the first non-zero order correction to the amplitude in the classical field.  Higher order calculation requires a higher order solution of the classical field from the Yang-Mills equation and also the inclusion of the higher order terms in the Lagrangian.  We defer the higher order consideration to future study.  

Nevertheless, the modified amplitude can be applied to the minijet cross section and provides interesting phenomenological implications.  After resumming the leading order correction, a suppression factor is introduced in the $gg \rightarrow gg$ amplitude.  This suppression factor depends on the $x$ values of the incident gluons of the minijet model.  The amplitude receives a stronger suppression as $x$ is smaller.  We calculated the minijet cross section in \cite{Cheung2011b} and found that this modified minijet model provides a satisfactory fit to the total cross section of $pp$ and $\pbar p$ for a wide range of energy and has a $(\ln s)^2$ behaviour at high energy.

\appendix*
\section{\label{sec:GI} Gauge invariance}
In this Appendix, we will use the functional method to derive a gauge invariant condition for the two-point Green's function of the propagator and use it to check gauge invariance.  We generalized the Slavnov-Taylor identity \cite{Slavnov1972,Taylor1971} to the presence of an non-zero background field.  

\subsection{Gauge transformation with background}
We start with the generating functional 
\begin{align}
  \mathcal{Z}[\eta;\rho]& = \int\mathcal{D}B \,\Delta(f(B)) \nonumber \\
\times &\exp\left\{i\int d^4x \left( \mathcal{L}_{0}(A+B,\rho) + \mathcal{L}_{GF} + \eta^a_\mu B^{a\,\mu}\right)\right\}
\end{align}
where $N$ is a normalization constant and the gauge fixing function $f$ is chosen as 
\begin{equation}
f^a(B) = \bar{D}^{ab}_\mu B^{b\,\mu}
\end{equation}
such that
\begin{equation}
  \mathcal{L}_{GF} = -\frac{1}{2} (\bar{D}_{ab}^\mu B^b_\mu)^2,
\end{equation}
Since the classical field $A$ is chosen to be a prescribed field, it does not transform.  As a result, $B$ takes all the burden of the gauge transformation.  The infinitesimal gauge transformation becomes
\begin{equation}
\left\{ \begin{array}{lll}
\rho^{a\,\mu} &\rightarrow & \rho'^{a\,\mu} = (\delta^{ab}-f^{abc}\alpha^c) \rho^{b\,\mu}\\
A^a_\mu &\rightarrow &A'^a_\mu = A^a_\mu\\
B^a_\mu &\rightarrow &B'^a_\mu = B^a_\mu + f^{abc}(A+B)^b_\mu \alpha^c +\frac{1}{g}\partial^\mu \alpha^a
\end{array} \right. \label{eq:app2:GT2}
\end{equation}
Hence, 
\begin{align}
  &\frac{\delta f^a(x)}{\delta \alpha^c(y)} \nonumber\\
&=\frac{1}{g}\left[ \delta^{ab}\square \delta(x-y)  \right.\nonumber \\
&+\left. gf^{adc} (2A^d_{x\,\mu} \partial^\mu +\partial^\mu A^d_{x\,\mu} + B^d_{x\,\mu} \partial^\mu +\partial^\mu B^d_{x\,\mu}) \delta(x-y)\right. \nonumber\\
&\left. +g^2 f^{aeb}f^{bdc} A^e_{x\, \mu} (A_x+B_x)^{e\,\mu}\delta(x-y)\right]. 
\end{align}
The subscript $x$ is an abbreviation of the argument of the function, e.g. $A_x = A(x)$.  
We define 
\begin{align}
M^{ac}(x,y)=&\bar{D}^{ab}_{x\,\mu} \left[ \partial^\mu \delta^{bc}+ gf^{bdc} (A_x+B_x)^{d \, \mu}\right] \delta(x-y) \label{eq:app2:M}
\end{align}
with an inverse $M^{-1\, cd}(x,y)$ satisfying
\begin{align}
\int dy &\bar{D}^{ab}_{x\,\mu} \left[ \partial^\mu \delta^{bc}\delta(x-y) \right. \nonumber \\
&+\left. gf^{bdc} (A_x+B_x)^{d \, \mu} \delta(x-y)\right] M^{-1\, cd}(y,z) \nonumber \\
& = \delta^{ad}\delta(x-z)
\end{align}
The determinant of the gauge fixing function with respect to the infinitesimal gauge transformation is
\begin{equation}
\Delta(f) = \det\left(\frac{\delta f^a(x)}{\delta \alpha^b(y)}\right) = \det(M)\det(\frac{1}{g}).
\end{equation}
The determinant of $1/g$ can be absorbed into the normalization constant $N$.  This determinant can also be written as a ghost term in the Largangian:
\begin{align}
\det(M) =& \int \mathcal{D}\bar{c}\mathcal{D}c \exp  \left\{-i\int \right.  d^4x \left[ \bar{c}_a \square c_a  \right. \nonumber \\
&+gf^{abc} \bar{c}_a (2A^b_\mu \partial^\mu +\partial^\mu A^b_\mu + B^b_\mu \partial^\mu +\partial^\mu B^b_\mu) c_c \nonumber \\
&+g^2 f^{aeb}f^{bdc} \bar{c}_a A^d_\mu (A+B)^{e\,\mu} c_c] 
\end{align}
Additional interactions between the ghost and the classical field $A$ are also introduced.  
The connected $n$-point Green's function of the quantum gluon is given by taking $n$-derivative with respect to the source $\eta$ then divided by $Z$ evaluating at $\eta=0$ 
\begin{equation}
G^{a_1,\dots a_n}_{\mu_1,\dots \mu_n}(x_1,\dots x_n) = \frac{(-i)^n}{Z} \frac{\delta^n Z} {\delta \eta^{\mu_1}_{a_1}(x_1)\dots\delta \eta^{\mu_n}_{a_n}(x_n)}\Bigg\vert_{\eta=0}
\end{equation}
\subsection{Slavnov-Taylor Identity}
As we mentioned in Section \ref{sec:3}, the observable has to be gauge invariant.  The transition amplitude depends explicitly on the Green's functions which itself depends on the gauge fixing.  Therefore, one needs to find out a set of conditions on the Green's functions to ensure gauge invariance.  This is equivalent to restricting the generating functional to be invariant under gauge transformation.  To do that, we transform the generating functional using Eq.(\ref{eq:app2:GT2}) such that
\begin{align}
&\mathcal{Z}[\eta;\rho] \rightarrow  
\mathcal{Z}'[\eta;\rho'] = N\int\mathcal{D}B' \,\Delta(f(B')) \, \nonumber \\
&\exp\left\{i\int d^4x \left( \mathcal{L}_{0}(A+B',\rho') + \mathcal{L}_{GF}(B') + \eta^a_\mu B'^{a\,\mu}\right)\right\}
\end{align}
Since $\mathcal{L}_0$ is gauge invariant, \[\mathcal{L}_0(A+B',\rho')=\mathcal{L}_0(A+B,\rho),\] and the Jacobian of the transformation on $\mathcal{D}B$ is $1$, \[\mathcal{D}B=\mathcal{D}B',\] the difference between $Z$ and $Z'$ is due to the change in gauge fixing term, the determinant $\Delta(f)$ and the source term $\eta B$.
\begin{widetext}
\begin{align}
  \mathcal{Z}'=& N\int\mathcal{D}B\,(\Delta + \delta\Delta) \exp \Biggl\{i\int d^4x \biggl( \mathcal{L}_{0} + \mathcal{L}_{GF}+\delta\mathcal{L}_{GF}+ \eta^a_\mu B^{a\,\mu} +\eta^a_\mu \delta B^{a\,\mu} \biggr) \Biggr\} \nonumber \\
\sim &\mathcal{Z} + N \int\mathcal{D}B\,\Delta \left[\frac{\delta\Delta}{\Delta}+
i\int d^4x \bigg(\delta \mathcal{L}_{GF}+\eta^a_\mu \delta B^{a\,\mu} \bigg) \right]\exp{ \Biggl\{i\int d^4x \biggl( \mathcal{L}_{0} + \mathcal{L}_{GF}+\eta^a_\mu B^{a\,\mu} \biggr) \Biggr\}}. \label{eq.GC1}
\end{align}
Gauge invariance requires that 
\begin{align}
0= \delta \mathcal{Z} =  N &\int\mathcal{D}B\,\Delta \left[\frac{\delta\Delta}{\Delta}+
i\int d^4x \bigg(\delta \mathcal{L}_{GF}+\eta^a_\mu \delta B^{a\,\mu} \bigg) \right]
\exp{ \Biggl\{i\int d^4x \biggl( \mathcal{L}_{0} + \mathcal{L}_{GF}+\eta^a_\mu B^{a\,\mu} \biggr) \Biggr\}}
\end{align}
\end{widetext}
for all $\alpha^a(x)$, where
\begin{align}
\delta\mathcal{L}_{GF}  
=-\left(\bar{D}^{ab}_\mu B^{b\, \mu} \right) \frac{1}{g} \int dy M^{ac}(x,y) \alpha^c(y) \label{eq.delta_GF}
\end{align}
and 
\begin{align}
 \eta^a_\mu \delta B^{a\,\mu}\nonumber 
=& \left[\eta^a_\mu f^{adc}(A+B)_x^{d\,\mu}-\frac{1}{g}\left(\partial^\mu \eta^a_\mu \right)\delta^{ac}\right] \alpha^c(x) \\
&+ \frac{1}{g} \partial^\mu \left( \eta^a_\mu \alpha^a  \right) \label{eq.delta_source}
\end{align}
From here on we will denote $\Delta(f)$ as the determinant of $M^{ab}(x,y)/g$.  After the transformation, $M$ becomes $M+\delta M$ and the determinant becomes, under gauge transformation (\ref{eq:app2:GT2}), 
\begin{align*}
\Delta' &= \det(1/g)\det (M + \delta M) \\
&=\det\Big( \bar{D}^{ab}_{x\,\mu} \big[ \partial^\mu \delta^{bc} + gf^{bdc} (A_x+B_x)^{d \, \mu} \\
& \quad +  gf^{bdc}f^{dmn} (A_x+B_x)^{m\,\mu} \alpha_x^n \\
& \quad + f^{bdc} (\partial_x^\mu \alpha^d_x) \big] \delta(x-y) \Big)
\end{align*}
Expanding $\Delta'$ to first order in $\delta M$, $\Delta' = \det(1/g)\det(M + \delta M) \approx \det(1/g)\Delta (1+Tr[\delta M M^{-1}])$.
One can identify $\frac{\delta \Delta}{\Delta}=Tr(\delta M M^{-1})$.
Explicitly, $M$ and $\delta M$ are 
\begin{align*}
M^{ac}_x  =& \bar{D}^{ab}_{x\,\mu} \left[ \partial^\mu \delta^{bc} + gf^{bdc} (A_x+B_x)^{d \, \mu} \right]\\
\delta M^{ac}=& \bar{D}^{ab}_{x\,\mu} \left[ gf^{bdc}f^{dmn} (A_x+B_x)^{m\,\mu} \alpha_x^n  \right. \\
& \left. + f^{bdc} (\partial_x^\mu \alpha^d_x)\right];
\end{align*}
therefore, the trace becomes\footnote{$Tr[f(x,y)] = \int dx dy \delta(x-y) f(x,y) = \int dx f(x,x)$}
\begin{align}
Tr&[\delta M M^{-1}] \nonumber \\
=&  Tr\left[\bar{D}^{ab}_{x\, \mu}\left[ gf^{bdc}f^{dmn} (A_x+B_x)^{m\,\mu} \alpha_x^n M^{-1\,c e}(x,y) \right. \right. \nonumber \\ 
& \left. + f^{bdc} (\partial_x^\mu \alpha^d_x) M^{-1\,c e}(x,y) \right]\delta^{ae} \delta(x-y) \nonumber \\
=&\int dx \, \left[ g f^{akl}f^{lcn} (\partial_x^\mu A_{x\,\mu}^k M^{-1\, c a}(x,x) ) \right. \nonumber  \\
& \left. + g^2 f^{akl}f^{ldc}f^{dmn}A^k_{x\,\mu} (A+B)^{m\, \mu}_x M^{-1 \, c a}(x,x)\right] \alpha^n_x \label{eq.delta_Delta}
\end{align}
By letting $\alpha^a_x = \int dy M^{ab}(x,y) \chi^b(y)$ and requiring the total contribution from eq.(\ref{eq.delta_GF}), (\ref{eq.delta_source}) and (\ref{eq.delta_Delta}) to the deviation of $Z$ to vanish, neglecting the total derivative term, we obtain the generalized Slavnov-Taylor identities 
\begin{widetext}
\begin{align}
\delta Z = 0 &= \Bigg\{ \int dx \, \left[ g f^{akl}f^{lcn} (\partial_x^\mu A_{x\,\mu}^k M^{-1\, c a}(x,x) ) \right. \nonumber  \left. + g^2 f^{akl}f^{ldc}f^{dmn}A^k_{x\,\mu} (A+B)^{m\, \mu}_x M^{-1 \, c a}(x,x)\right] M^{-1\, ne}(x,y)\nonumber\\
&\qquad - \frac{1}{g}\bar{D}^{eb}_\mu B^{b\, \mu}_y +\int dx \,\left[\eta^a_\mu f^{adc}(A+B)_x^{d\,\mu}-\frac{1}{g}\left(\partial^\mu \eta^a_\mu \right)\delta^{ac}\right] M^{-1\, ce}(x,y)\Bigg\} \mathcal{Z} \label{eq:app2:GST0}
\end{align}
\end{widetext}
The equation is an abbreviation of $\{\dots\}\mathcal{Z}=\int\mathcal{D}B \{\dots\} e^{i\int dx \mathcal{L}_0+\mathcal{L}_{GF}+\eta B}$.

To the leading order, the first two terms in eq. (\ref{eq:app2:GST0}) are neglected because they are at least one $g$ order higher than the last two.  Differentiating the last two terms with respect to $\eta^c_\nu(z)$ and setting all $\eta =0$, we obtain
\begin{align}
&0 = \Big\{ \bar{D}^{eb}_{\mu\,y} B^{b\, \mu}_y B^{c\,\nu}_z \nonumber \\
&+i \left[g f^{cdb}(A+B)_z^{d\,\nu}+\partial^\nu \delta^{cb}\right] M^{-1}_{be}(z,y)\Big\} \mathcal{Z}\big|_{\eta=0} 
\end{align}
By taking a covariant derivative $\bar{D}^{e'c}_{z\,\nu}$, the second term becomes delta functions according to eq. (\ref{eq:app2:M}) leading to  
\begin{equation}
\Big\{ \bar{D}^{eb}_{\mu\,y} \bar{D}^{dc}_{\nu\,z} B^{b\, \mu}_y B^{c\,\nu}_z + i \delta^{ed}\delta(z-y) \Big\} \mathcal{Z}\big|_{\eta=0} =0 \label{eq:app2:GST1}
\end{equation}
By identifying the connected two-point Green's function as \[G^{bc}_{\mu\nu}(x,y) = \frac{1}{\mathcal{Z}}\big\{ B^{b\, \mu}_x B^{c\,\nu}_y\big\}\mathcal{Z}\big|_{\eta=0}\] and taking the Gaussian average of eq. (\ref{eq:app2:GST1}), it leads to
\begin{equation}
 \langle\bar{D}^{ac\,\mu}_{x}\bar{D}^{bd\,\nu}_{y} G_{\mu\nu}^{cd}(x,y;\rho) \rangle = -i\delta^{ab}\delta^4(x-y) \label{eq:app2:GST2}
\end{equation}
This condition is analogous to the transverality condition of gluon Green's function in Slavnov's original derivation (eq. (22) and (23) in \cite{Slavnov1972}).  One can check the propagator derived in Section 5 satisfies eq. (\ref{eq:app2:GST2})

\subsection{Gauge invariance of the modified propagator}
We will show the Green's function 
\begin{equation}
G_0{}^{ab}_{\mu\nu} + G^+_{11}{}^{ab}_{\mu\nu}, \label{eq:app2:GF}
\end{equation}
obtained in eq. (\ref{eq:4:G11}), to the leading order in $g^2$ and $A^2$ satisfies the generalized Slavnov-Taylor identity, eq. (\ref{eq:app2:GST2}).  We ignore the term containing the interaction of the second type, $\mathcal{L}_{int_2}$, since it does not contribute to the scattering amplitude as discussed in Section \ref{sec:4}.  For simplicity, only one proton will be considered.  Since the contributions of the cross terms involving two colliding sources vanishes after the Gaussian average, the proof can be easily generalized to the case of two protons.

In the LHS of eq. (\ref{eq:app2:GST2}), the Gaussian average is taken after acting the covariant derivative on the Green's function (GF).  When the term linear to $A$ in the covariant derivative multiplies with the first order term in $A$ of the GF, the Gaussian averaged product will have non-zero contribution.  Therefore, besides the Gaussian averaged terms (eq. \ref{eq:app2:GF}) obtained in Section \ref{sec:4}, one also needs to include the diagram that has only one interaction with the classical field in $G$ of the LHS of eq. (\ref{eq:app2:GST2}) before taking the average.  

\subsection{Term linear in $\mathcal{L}_{int_1}$}
\begin{figure}[t]
\centering
\includegraphics[width=0.6\linewidth]{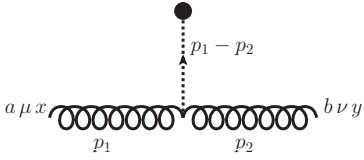}
\caption{Schematic Feynman diagram for gluon GF with single interaction with $A$.}
\label{fig:app2:1}
\end{figure}

We have calculated the terms of $O(A^2)$ in Section \ref{sec:4}.  To check gauge invariance, we also need another term that vanish only after taking the Guassian average.  This is the term that the gluon interacts with $\mathcal{L}_{int_1}$ once.  Diagrammatically, we consider a gluon enters from the left with momentum $p_1$ then interacts at $z$ and leaves with momentum $p_2$ as shown in Fig. (\ref{fig:app2:1}).
The GF is 
\begin{align}
&G^I{}_{\mu\nu}^{ab}(x,y) = i \langle T{B^a_\mu(x) \int dz \mathcal{L}_{int_1}(z) B^b_\nu(y)}\rangle \nonumber \\
=& -i gf^{cmd}\int dz  \langle T{B^a_\mu(x)g^{\alpha\beta} B^c_\alpha (z) A^{c\,+}\partial^- B^d_\beta(z) B^b_\nu(y)}\rangle.
\end{align}
There are two way to contract the fields.  The sum of the two contributions are
\begin{eqnarray}
G^I{}_{\mu\nu}^{ab}(x,y) = &&g\, g_{\mu\nu} f^{abc}\int d^4z \frac{d^4p_1 d^4p_2}{(2\pi)^8} \frac{(p_1^-+p_2^-)}{p_1^2 p_2^2} \nonumber \\ 
&&A^{c\,+}(z) e^{-i p_1(x-z)- i p_2(z-y)}. \label{eq:app2:GI}
\end{eqnarray}

\subsection{Checking gauge invariant}
Let us rewrite the LHS of eq. (\ref{eq:app2:GST2}) according to the order of $A$ as
\begin{align}
& \langle \bar{D}^{ac\,\mu}_{x}\bar{D}^{bd\,\nu}_{y} G_{\mu\nu}^{ab}(x,y) \rangle \nonumber \\
&= \langle \partial_{x}^\mu\partial_{y}^\nu G_{0}{}^{ab}_{\mu\nu}(x,y)\rangle \nonumber \\
&+ \langle \partial_{x}^\mu gf^{bcd}A^{c\,\nu}_y G^I{}_{\mu\nu}^{ad}(x,y) + \partial_{y}^\nu gf^{acd}A^{c\,\nu}_x G^I{}_{\mu\nu}^{db}(x,y) \rangle \nonumber \\
&+ \langle \partial_{x}^\mu\partial_{y}^\nu G_{11}^{+}{}_{\mu\nu}^{ab}(x,y) \rangle. \label{eq:app2:GST}
\end{align}
The first term is the derivative on a bare Feynman propagator.  The Gaussian average provides no effect to this term.  Straight forward evaluation gives
\begin{align*}
&\delta^{ab}g_{\mu\nu} \partial_x^\mu \partial_y^\nu \int \frac{d^4 q}{(2\pi)^4}  \frac{-i}{q^2} e^{-i q (x-y)} \\
=& \delta^{ab}\int \frac{d^4 q}{(2\pi)^4} (-i) e^{-i q (x-y)} \\
=& -i \delta^{ab} \delta(x-y),
\end{align*}
which is the RHS of eq. (\ref{eq:app2:GST2}).  The second term 
\begin{align*}
&\langle \partial_{x}^\mu gf^{bcd}A^{c\,\nu}(y) G^I{}_{\mu\nu}^{ad}(x,y) \\
&=g^2 f^{bcd}f^{ade} \langle A^{c\,+}(y)A^{e\,+}(z) \rangle \nonumber \\
& \quad \partial_x^- \int d^4z \frac{d^4p_1 d^4p_2}{(2\pi)^8} \frac{(p_1^-+p_2^-)}{p_1^2 p_2^2}  e^{-i p_1(x-z)- i p_2(z-y)}.
\end{align*}
Using eq. (\ref{eq:4:AA2}) to evaluate the Gaussian average, we have
\begin{align*}
& \langle \partial_{x}^\mu gf^{bcd}A^{c\,\nu}(y) G^I{}_{\mu\nu}^{ad}(x,y) \\
=& \delta^{ab} i\frac{4 \alpha_s N_c g^2}{(2\pi)^6}  (p_2^-)^2  \int d^4p_2 \frac{1}{p_2^2} \int\frac{dp_1^+}{2p_1^+p_2^- -p_{1\perp}^2 +i\epsilon} \nonumber \\
&\int \frac{d^2 k_{\perp}}{k_\perp ^4}\bar{\mu}_x(k_\perp^2)e^{-i p_2 (x-y)}.
\end{align*}
Identifying the integrals of $p_1$ and $k_\perp$ as the same as of eq. (\ref{eq:4:dp+}) and (\ref{eq:4:dk}), and by changing $p_2$ to $q$, we finally have 
\begin{align}
&\langle \partial_{x}^\mu gf^{bcd}A^{c\,\nu}(y) G^I{}_{\mu\nu}^{ad}(x,y) \nonumber \\
=& \delta^{ab} \frac{\alpha_s N_c}{2\pi}   \left(\theta(q^-)-\theta(-q^-)\right) I(Q^2,x) \nonumber \\ 
&\int \frac{d^4q}{(2\pi)^4}\frac{ q^-}{q^2}  e^{-i q (x-y)}.
\end{align}
The third term of eq. (\ref{eq:app2:GST}) term has the same contribution as the first term, so the sum of the two terms is 
\begin{align}
 \delta^{ab} \frac{N_c g^2}{4\pi^2}   \left(\theta(q^-)-\theta(-q^-)\right) I(Q^2,x) \int \frac{d^4q}{(2\pi)^4}\frac{ q^-}{q^2}  e^{-i q (x-y)} \label{eq:app2:12}
\end{align}
Let's recall the last term in eq. (\ref{eq:app2:GST}) as the coordinate space representation of eq. (\ref{eq:4:G11}), 
\begin{eqnarray}
G^+_{11}{}^{ab}_{\mu\nu}(x,y)=&&-g_{\mu\nu}\delta^{ab} \frac{\alpha_s N_c}{\pi} I(Q^2,x) (\theta(q^-)-\theta(-q^-)) \nonumber \\
&&\int \frac{d^4q}{(2\pi)^4}\frac{ q^-}{q^4}  e^{-i q (x-y)}.
\end{eqnarray}
Applying the derivatives $\partial_x^\mu \partial_y^\nu$ provides $-iq^\mu iq^\nu$ which contracts with $g_{\mu\nu}$ gives $q^2$ canceling one of the $q^2$ in the denominator.  Therefore, we have 
\begin{align}
\partial_{x}^\mu&\partial_{y}^\nu \langle G_{11}^{+}{}_{\mu\nu}^{ab}(x,y) \rangle \nonumber \\
=&-\delta^{ab} \frac{\alpha_s N_c}{\pi} I(Q^2,x)(\theta(q^-)-\theta(-q^-)) \nonumber \\ 
&\int\frac{d^4q}{(2\pi)^4}\frac{ q^-}{q^2}  e^{-i q (x-y)},
\end{align}
which exactly cancels the sum of the two previous terms in (\ref{eq:app2:12}).  Therefore, the generalized Slavnov-Taylor identity for two point GF is satisfied, to the leading order in $A^2$ and $g^2$, by the modified gluon propagator obtained in Section \ref{sec:4}.

So far, we have ignored the GF with the interaction term $\mathcal{L}_{int_2} \sim gB_\mu \bar{F}^{\mu\nu}B^\nu \sim g B^- \partial_i A^+ B^i$ in eq. (\ref{eq:4:Lint2}) for the practical reason that these corrections do not contribute to the scattering amplitude.  However, the authors in \cite{Wang02} indicated that the typical scale of the background field tensor $\bar{F}$ goes like $g A^2$.  Therefore, one can consider $\mathcal{L}_{int_2}$ as a higher order term compared to $\mathcal{L}_{int1}$.  Ignoring $\mathcal{L}_{int_2}$ can be valid even at the level of order counting.  This justifies our result.  

\begin{acknowledgments}
We would like to thank Prof. E.C.G. Sudarshan and Prof. Duane Dicus for helpful discussions.  
\end{acknowledgments}

\bibliography{prd_2011_oct}

\end{document}